\documentclass[fleqn,usenatbib]{mnras}

\usepackage[T1]{fontenc}
\usepackage{ae,aecompl}
\usepackage[normalem]{ulem}

\usepackage{graphicx}	
\usepackage{amsmath}	
\usepackage{amssymb}	

\graphicspath{{./../files/}}

\title[Fomalhaut]{1.3 mm ALMA Observations of the Fomalhaut Debris System}

\author[J. A. White et al.]{
Jacob Aaron White$^{1}$\thanks{E-mail: jawhite@astro.ubc.ca},
A. C. Boley$^{1}$,
W. R. F. Dent$^{2}$,
E. B. Ford$^{3,4,5,6}$,
S. Corder$^{7}$
\\
$^{1}$Department of Physics and Astronomy, 6224 Agricultural Road, Vancouver, BC V6T 1Z1\\
$^{2}$ALMA SCO, Alonso de C\'ordova 3107, Vitacura, Santiago, Chile\\
$^{3}$Center for Exoplanets and Habitable Worlds, The Pennsylvania State University, University Park, PA 16802\\
$^{4}$Center for Astrostatistics, The Pennsylvania State University, University Park, PA 16802\\
$^{5}$Institute for CyberScience, The Pennsylvania State University, University Park, PA 16802\\
$^{6}$Department of Astronomy \& Astrophysics, The Pennsylvania State University, University Park, PA 16802\\
$^{7}$North American ALMA Science Center, NRAO, 520 Edgemont Road, Charlottesville, VA, 22903, USA
}

\date{Accepted XXX. Received YYY; in original form ZZZ}

\pubyear{2016}

\begin{document}
\label{firstpage}
\pagerange{\pageref{firstpage}--\pageref{lastpage}}
\maketitle

\begin{abstract}
We present ALMA Band 6 observations (1.3 mm/233 GHz) of Fomalhaut and its debris disc. The observations achieve  a sensitivity of 17 $\mu$Jy and a resolution of 0.28 arcsec (2.1 au at a distance of 7.66 pc), which are the highest resolution observations to date of the millimetre grains in Fomalhaut's main debris ring. The ring is tightly constrained to $139^{+2}_{-3}$ au  with a FWHM of $13\pm3$ au, following a Gaussian profile. The millimetre spectral index is constrained to $\alpha_{mm} = -2.73\pm0.13$. We explore fitting debris disc models in the image plane, as well as fitting models using visibility data directly. The results are compared and the potential advantages/disadvantages of each approach are discussed. 

The detected central emission is indistinguishable from a point source, with a most probable flux of $0.90\pm 0.12$ mJy  (including calibration uncertainties). This implies that any inner debris structure, as was inferred from far-Infrared observations, must contribute little to the total central emission. Moreover, the stellar flux is less than 70\% of that predicted by extrapolating a black body from the constrained stellar photosphere temperature. This result emphasizes that unresolved inner debris components cannot be fully characterized until the behaviour of the host star's intrinsic stellar emission at millimetre wavelengths is properly understood.

\end{abstract}




\section{Introduction}

Fomalhaut is one of the Sun's closest stellar neighbours and has been the target of numerous studies at multiple wavelengths (see Table \ref{flux_list} for a select list of observations). 
Located at a distance of $7.66 \pm 0.04$ pc \citep{hipparcos}, this $200-440$ Myr old A3V star \citep{difolco, mamajek} has a bright, eccentric debris ring at stellar separation of about 140 au, which can serve as an important testbed for debris disc evolution models and potentially planet-disc interactions. 

Despite extensive study, outstanding issues remain in characterizing Fomalhaut's debris system at millimetre (mm) wavelengths. Two of such issues are (1) determining whether there is a debris component that is interior to the 140 au main ring, and (2) constraining millimetre flux densities of the outer ring, thus determining the millimetre spectral index.  Both of these are directly related to understanding the evolution of the debris system itself, as well as using the debris to constrain the structure of Fomalhaut's putative planetary system. 

The possible presence of a warm, inner debris disc was first identified by \citet{stapelfeldt} after unresolved excess, compared with the expected stellar emission, was found at 24 $\mu$m using \textit{Spitzer} data. To better improve the flux uncertainty, a re-reduction of the \textit{Spitzer} data by \citet{su16} finds an excess of $0.64 \pm 0.13$ Jy ($\sim20\%$ over a Kurucz model atmosphere). Using archival VLTI data, \citet{absil} report an excess of $0.88\% \pm 0.12\%$ over the stellar photosphere in the K band, although they are unable to distinguish between a point source and an extended source for the central emission. \citet{acke} fit a 3 component model consisting of a point source, an inner disc, and an outer disc to \textit{Herschel} data and find an unresolved excess of $0.17 \pm 0.02$ Jy, or about 50\% over the expected stellar flux.  In contrast, ALMA $870~\mu$m observations \citep{su16} find a total flux about 35\% \textit{lower} than what is expected from the stellar photosphere and detect no extended structures within 0.2 arcsec ($\sim15$ au) of the central emission.

For the main ring, the grain size distribution, $q$, can be used as a tracer of the collisional processes that are present in the late stages of planet formation \citep{dohnanyi}. The size distribution can be further influenced by gravitational interactions with a nearby massive planet or by self-stirring within the disc \citep{wyatt08, mustill}. Smaller $\mu$m sized grains are subject to strong interactions with radiation pressure, while the larger mm grains will be better tracers of the parent body distribution. The distribution of mm grains in Fomalhaut's outer disc can therefore be used to test the collisional models of planetesimals and the dynamical state of the system \citep{vandenbussche, ricci}. Previous constraints find $q$ to be between 3.4 and 4 \citep{ricci, pan}; a more precise measurement is needed to describe the dynamical state of the Fomalhaut debris disc.

Finally, we note that there is a well-known scattered light feature located NW of the star, just inside the outer debris ring \citep{kalas2008}.  While the nature of the source is debated, it could be directly related to the putative planet Fomalhaut b, or it could also be a byproduct of collisional processes \citep{lawler}. Observations at infrared wavelengths have failed to detect a planet \citep{marengo, janson} in the system. Regardless, the significant eccentricity of the debris ring would be consistent with perturbations from planets  \citep{kalas,quillen2006}, although hydrodynamic processes could also play a role if there is sufficient gas \citep{lyra}. If the feature and/or the ring's eccentricity are related to a planetary system, then Fomalhaut would be an ideal system for studying planet disc-interactions.

In this paper we present 1.3 mm ALMA data that provide the highest resolution (0.28 arcsec) observations to date of the outer mm-debris disc and allow us to resolve the regions immediately around the star to within about 2 au.  The observations are thus able to address whether there is indeed excess over the stellar emission, constraining the presence of an inner debris disc or ring. Section 2 is an overview of the observations and data reduction. Section 3 details the image-plane and visibility modelling of the debris disc and central emission. Section 4 shows a fit to the SED of the host star, calculates the grain size distribution, and discusses the central emission. Section 5 summarizes the results.

\section{Observations}

The data were acquired as part of the ALMA cycle 2 campaign (project ID 2013.1.00486.S). Observations were made in three execution blocks (EBs) taking place between 2015 June 11$^{\rm th}$  and 2015 September 21$^{\rm st}$. The average integration time was 1.08 hr.  A 34-antenna configuration was used; the longest baseline was 1.1 km.  Observations were centred on Fomalhaut using J2000 coordinates RA = 22 h 57 min 39.44 s  and $\delta = -29^{\circ} 37' 22.64''$. The observations were taken in band 6 (at $\sim233$ GHz) with the correlator setup using the Time Division Mode (TDM) and dual polarization. Four different spectral windows were used with 2 GHz bandpasses at rest frequency centres of 224, 226, 240, and 242 GHz.  Each spectral window had 128 channels with a corresponding channel width of 15.625 MHz.

Ceres and quasar J2258-2758 were used for absolute flux and bandpass calibration, respectively. Atmospheric variations at each antenna were monitored continuously using the water vapor radiometer (WVR). Data were reduced using the Common Astronomy Software Applications ({\scriptsize CASA}) package \citep{casa_reference}. The data reduction in {\scriptsize CASA} included WVR calibration; system temperature corrections; and bandpass, flux, and phase calibrations with Ceres and quasar J2258-2758. The size of the synthetic beam is $0.329 \times 0.234$  arcsec at a position angle of 83.6$^{\circ}$. The beam corresponds to $\sim 2.1$ au at the system distance of 7.66 pc. The FWHM of the primary beam is 27.3 arcsec at the wavelength of the observations.

\begin{figure}
\centering
\includegraphics[width=0.5\textwidth]{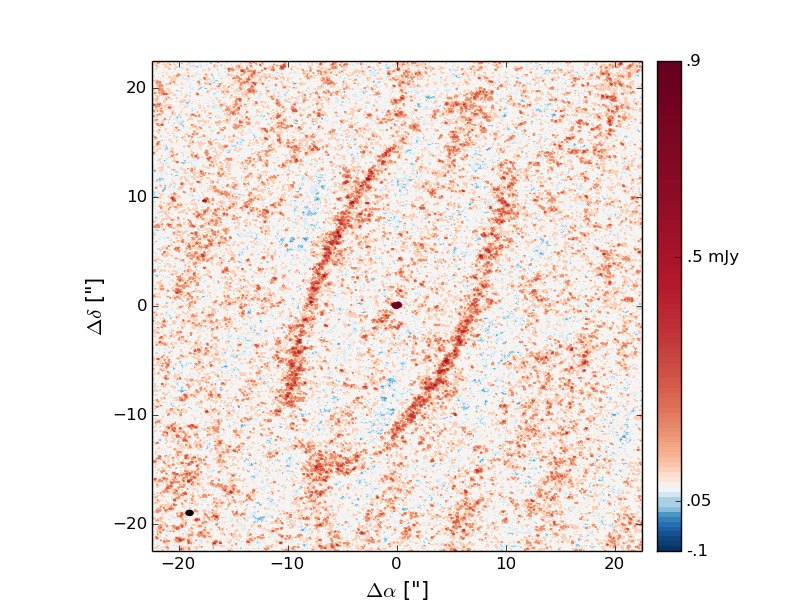}
\caption{CLEANed data of the Fomalhaut system. The synthetic beam is given by the black ellipse in the bottom left of the image.  Coordinates are given as offset from the phase centre. North is up and East is to the left.
\label{fig1}}
\end{figure}

The CLEANed image is shown in Fig.\,\ref{fig1}. This 233 GHz continuum image was produced by {\scriptsize CASA}'s \textit{CLEAN} algorithm  using threshold of $\frac{1}{2}~ \sigma_{\rm RMS}$ and natural weighting. The average wavelength across the frequency range is 1296 $\mu$m. The central emission and the main debris ring are clearly detected, but no extended structure is found interior to the main ring. To analyze the system further, we fit the data with multi-component models. 

\section{Debris Disc Modelling}

We constrain the morphology of the Fomalhaut debris system by conducting a search through model parameters using a Markov Chain Monte Carlo (MCMC) approach.  For any given model, we represent the system using a point source and a circular ring that has a Gaussian radial profile for the debris.  The ring's centre is allowed to be offset from the star's position, approximating a low-eccentricity ellipse.  We do not include an additional inner debris disc/ring here.  The choice of a Gaussian radial profile for the debris's spatial distribution is motivated in part by the system's similarities to the Solar System's Kuiper Belt \citep{kavelaars, boley}.  

In our analysis, we fit the following parameters: debris disc centre, disc radial width, system inclination relative to the observer, disc position angle, X-offset and Y-offset, disc flux, and central emission flux. The X and Y-offsets are the projected angular distances (measured in arcsec) of the central emission relative to the geometric centre of the ring. Models are generated using the same approach as in \citet{white} and assume a flat prior distribution. In generating a model, particles are randomly distributed within the debris ring, according to the given profile. The local grain temperature is derived assuming thermal equilibrium with the star and is azimuthally symmetric around the center of the disk. We assume that all the grains in the disc are perfect radiators in radiative equilibrium with the star.  The model is then rotated to a trial sky position and projected onto a grid to get the unscaled brightness distribution. The final brightness distribution of the ring is then set by demanding that the system has a total flux density that is consistent with the given MCMC link. The central emission of the system (i.e., star and any stellar excess) is modelled as a point source at the phase centre of the observations, consistent with the lack of extended emission seen in Fig.\,\ref{fig1}.

For each model, we assume that the vertical profile of the debris is Gaussian with a width  of $1^\circ$, as viewed from ring's centre \citep[see, e.g.,][]{boley}.  In principle, the ring opening angle can be constrained, albeit very approximately, by effectively comparing the width of the ring at the ansae with the width of the ring close to quadrature.   However, in the 1300 $\mu$m observations presented here, the ring ansae are too close to the edges of the primary beam to produce  meaningful, independent results. 

We use two different approaches in fitting models to the data. In the first method, following \cite{booth}, we fit the data in the image plane by producing dirty images for each model (discussed more below).  For the second and more standard method, we use each model to predict visibilities and then compare those results with the actual visibility data. In both cases, parameter space is explored using a random walk directed by a Metropolis-Hastings MCMC  \citep[for a review of MCMC see][]{ford}. For each new trial, two model parameters are randomly chosen and then updated by drawing a Gaussian random parameter centred on the current model (state $i$). The acceptance probability for the new trial model (state $i+1$) is given by 
\begin{equation}
\alpha =\rm{min}(e^{\frac{1}{2}\left(\chi^{2}_{i}-\chi^2_{i+1}\right)},1).
\end{equation}
The $\chi^{2}$ is different depending whether the fit is done in the image or visibility plane; the corresponding forms are described below in Sections \ref{sec:imageplane} and \ref{sec:visibilityplane}.  
If for a given $\chi^{2}$, $\alpha$ is greater than a random number drawn from a uniform [0,1] distribution, then the new model is accepted and recorded in the Markov chain. 
If the model is rejected, then the previous model is used again and re-recorded. The resulting chains are thinned by a factor of 10 (i.e. every 10$^{th}$ link is used) and used to determine the posterior distributions.  The thinned chains are also checked for convergence using a k-lag autocorrelation function (ACF). The ACF tests how well the sample is mixed by comparing a given parameter x$_{i}$ to a parameter further in the chain, x$_{i+k}$. The less the chain is autocorrelated, the lower the lag needed, k, for the ACF to drop to near 0.

\subsection{Image Plane\label{sec:imageplane}}

Given the high resolving power of ALMA, accurate modelling of interferometric data in the image plane is becoming more plausible. This has some advantages over visibility modelling, since it can be considerably faster and less complex to model in the image plane. Visibilities do not need to be calculated for each model, and the number of calls to, e.g.,  {\scriptsize CASA} is greatly reduced.

We carry out the MCMC modelling in the image plane based on the method described in \citet{booth}.  A trial two-component model is constructed using a debris ring and central emission (as described above). The model is attenuated by the primary beam and then convolved with the synthetic beam (the sampling function). The primary and synthetic beams are obtained from creating a ``dirty" image in {\scriptsize CASA} using the \textit{CLEAN} task (with zero iterations) and then exporting to a more manageable format via the \textit{exportfits} task.  The resulting model dirty image is then compared with the actual dirty image from the observations by using a $\chi^2$ statistic of the form
\begin{equation}
\chi^{2}_{i} = \sum{ \frac{(I - M_{i})^{2} }{\sigma^{2} } }
\end{equation}
where $I$ are the data from the dirty image, M$_{i}$ is the current model, and $\sigma = 374~\mu$Jy is the $\sigma_{\rm RMS}$ of the dirty image multiplied by the beam size in pixels \citep[see][]{booth}. The summation is over all pixels in the image. If $\alpha$ is greater than a random number drawn from uniform [0,1] distribution, the new model is accepted and recorded in the Markov chain. If the model is rejected, then the previous model is used again and re-recorded. 

\begin{figure*}
\centering
\includegraphics[width=\textwidth]{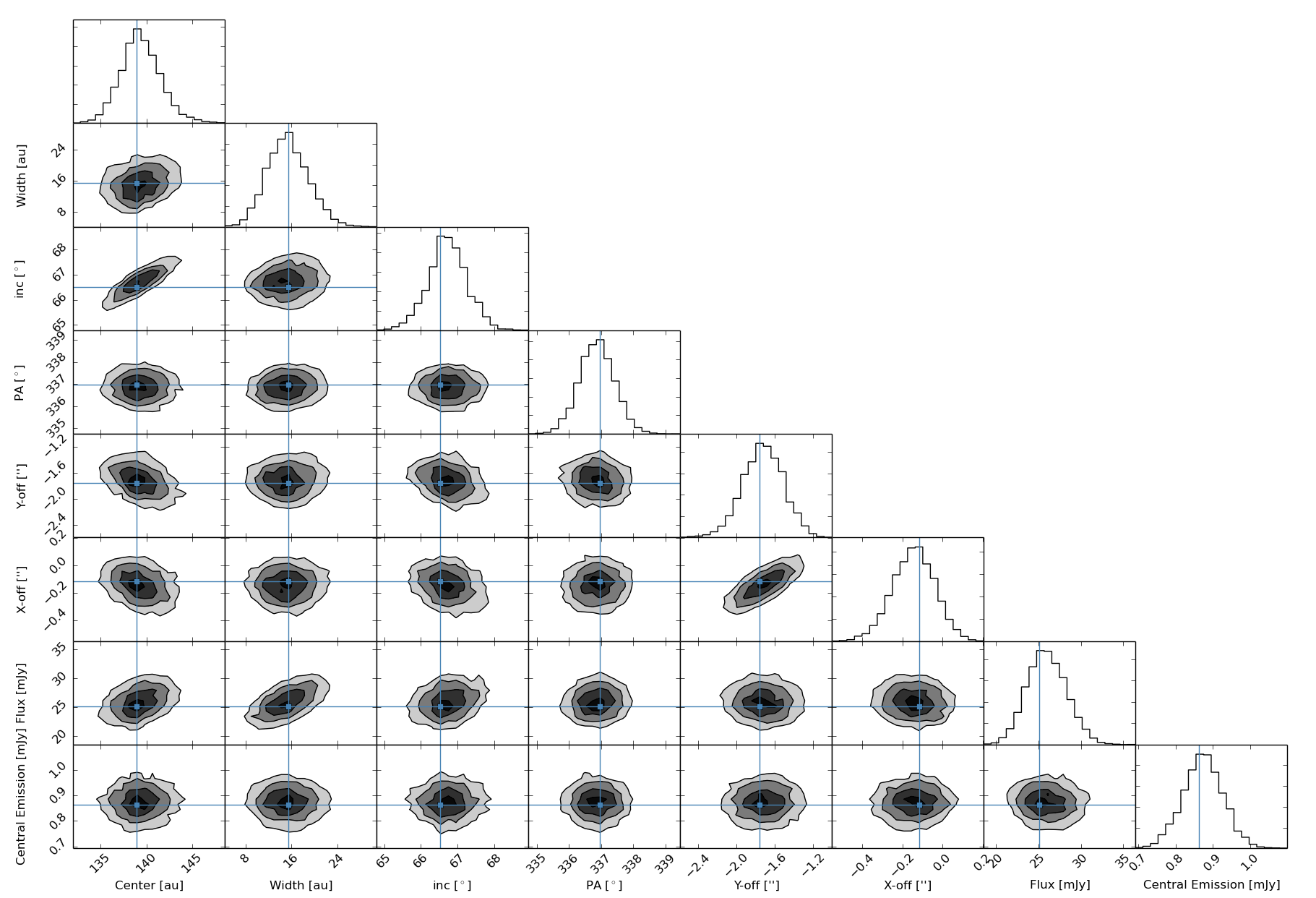}
\caption{MCMC parameter posterior distributions from the image plane fit. The blue points represent the most probable values.  The ``width'' is the  FWHM of the ring's radial Gaussian profile.
\label{mcmc_img}}
\end{figure*}

The MCMC routine is run with 3 separate chains for a total of 100000 links (minus about 1000 each for burn-in). The acceptance rate for the chain is $\sim 23\%$. The ACF becomes negligible for lags of 50 or less for all thinned chains. The 3 chains converge on similar parameters, and the distributions are combined to give the resulting posterior distributions in Fig. \ref{mcmc_img}.

The blue points correspond to the values of highest probability. The most probable parameters (i.e., the mode of the distributions) are given in Table \ref{mcmc_img_par}.  Uncertainties are given by a $95\%$ credible interval around the most probable value.

\begin{table}
\caption{Summary of image plane MCMC Results with $95\%$ Credible Range. The ``width'' is the  FWHM of the ring's radial Gaussian profile.}
\centering 
\begin{tabular}{c | c | c} 
\hline\hline 

    Parameter & Most Probable & $95\%$ Credible Range \\
	\hline
	Centre [au]     & $139$  & $[135, 144]$ \\
	Width [au]   & $15$ & $[9, 22]$ \\
	Inclination [$^{\circ}$]   & $66.5$ & $[65.7, 67.8]$   \\
	Position Angle [$^{\circ}$]& $337.0$  & $[338.0, 335.8]$     \\
    X Offset [arcsec] & $-0.12$ & $[-0.36, 0.06]$ \\
    Y Offset [arcsec] & $-1.74$ &  $[-2.16, -1.32]$ \\
	Disc Flux [mJy] & $26.3$ & $[21.6, 30.8]$ \\
	Central Emission [mJy] & $0.89$ & $[0.78, 0.98]$ \

\label{mcmc_img_par}
\end{tabular}
\end{table}

\begin{figure*}
\centering
\includegraphics[width=\textwidth]{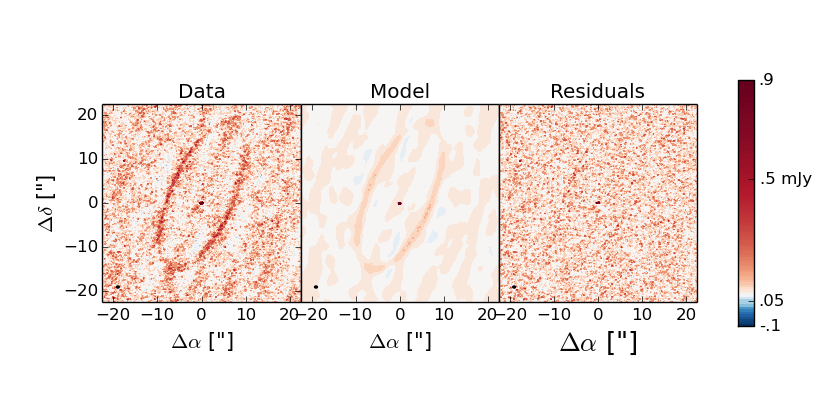}
\caption{\textbf{Left:} Dirty Image of the Fomalhaut system used in image plane model fitting. The synthetic beam is given by the black ellipse in the bottom left of the image. \textbf{Middle:} MCMC constrained best fit model of the system. Image is convolved with the synthetic beam and attenuated with the primary beam of the observations. \textbf{Right:} The data minus model residuals from the best fit model. The residuals are consistent with $\sim\sigma_{\rm RMS}$ of the observations. In all images North is up and East is to the left. The apparent excess in the center of the image is an artefact due to gridding effects in generating the image. There is also a slight residual feature in the location of the ring due to the image-plane fit not recovering the total flux of the disc (see Section 3.3). 
\label{img_resid}}
\end{figure*}

\subsection{Visibility Plane\label{sec:visibilityplane}}

To model the ALMA observations of Fomalhaut using the visibility plane, we first construct a trial, two-component sky model image of the debris ring and central emission (as already discussed) for the representative frequency of each spectral window. The sky model is then loaded into {\scriptsize CASA} and used to ``predict'' the visibilities that the model would have for the actual array configuration and (u,v) coordinates using the tasks \textit{setvp} and \textit{predict}. Each position in (u,v)-space has a real and imaginary component, and a corresponding weight. The weights for ALMA visibilties are 
\begin{equation}
WT = WEIGHT_{i,j} = \frac{w_{i} w_{j}}{\sigma^{2}_{i,j}},
\end{equation}
where $w_{i}$ and $w_{j}$ are antenna-based calibration factors derived by the {\scriptsize CASA} task \textit{applycal} during the data reduction process, and 
\begin{equation}
\sigma = \frac{1}{\sqrt{2 \Delta \nu \Delta t}},
\end{equation}
where $\Delta\nu$ and $\Delta t$ are the channel bandwidth and integration time. A $\chi^{2}$ is then calculated for each model visibility via
\begin{equation}
\chi^{2} = \sum_{i,j} (R_{i,j}^{D}-R_{i,j}^{M})^2 WT + (I_{i,j}^{D}-I_{i,j}^{M})^2 WT,
\end{equation}
where $R^{D}$, $R^{M}$ are the real components of the data and model visibilities; $I^{D}$, $I^{M}$ are the imaginary components of the data and model visibilities; and WT is the weights as given above. If $\alpha$ is greater than a random number drawn from uniform [0,1] distribution, the new model is accepted and recorded in the Markov chain. If the model is rejected, then the previous model is used again and re-recorded. 

\begin{figure*}
\centering
\includegraphics[width=\textwidth]{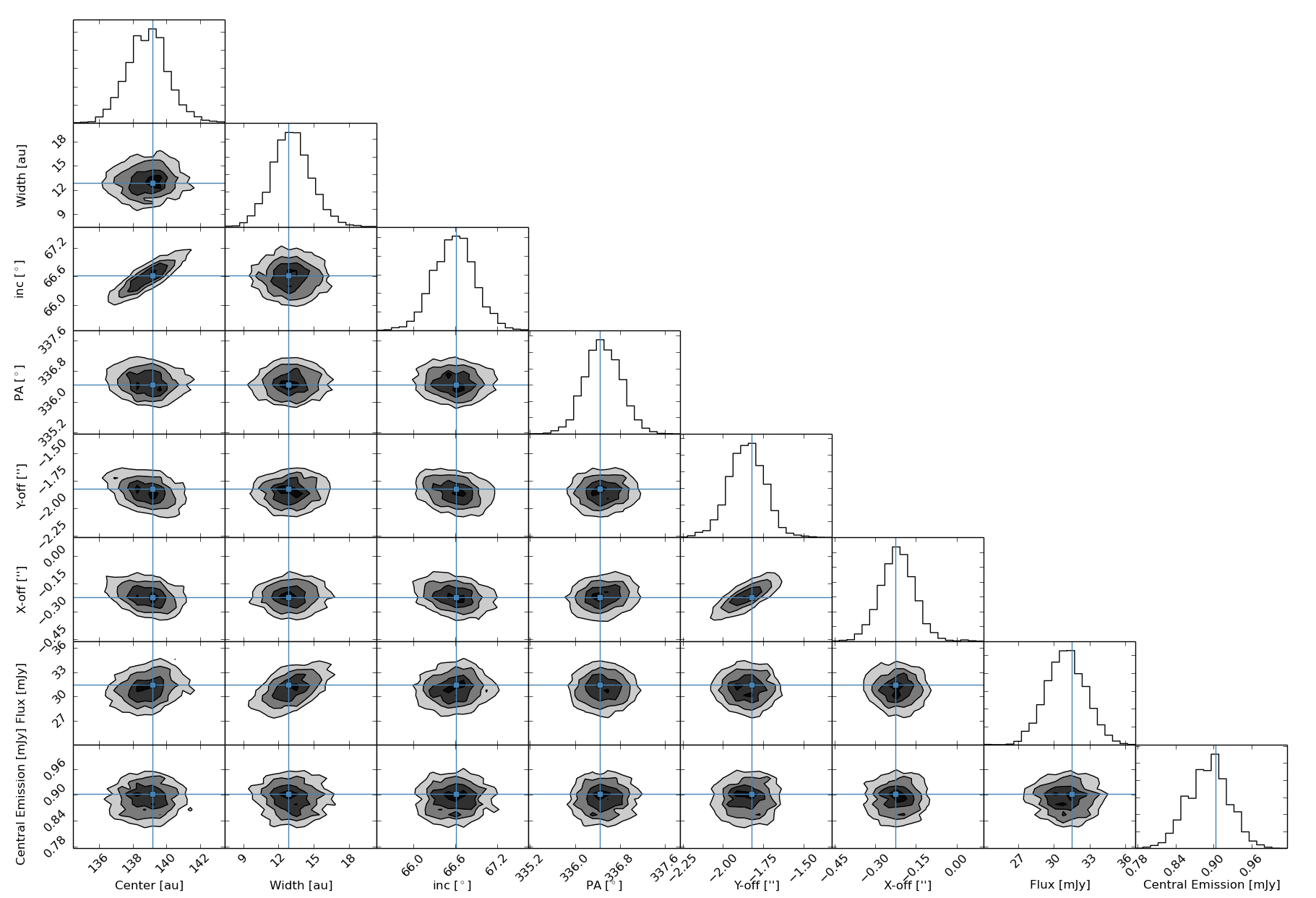}
\caption{MCMC parameter posterior distributions from the visibility plane fit. The blue points represent the most probable values. As before, the width is the FWHM of the ring's radial profile. 
\label{mcmc_vis}}
\end{figure*}

The MCMC routine is run with 10 separate chains for a total of 100000 links (minus about 1000 each for burn-in). The acceptance rate for the chain is $\sim 26\%$. The ACF becomes negligible for lags of 50 or less for all thinned chains The chains converge on similar parameters, and the distributions are combined to give the resulting posterior distributions in Fig. \ref{mcmc_vis}. The blue points correspond to the values of highest probability. The most probable parameters (i.e., the mode of the distributions) are given in Table \ref{mcmc_vis_par}.  Uncertainties are given by a $95\%$ credible interval around the most probable value.

\begin{figure*}
\centering
\includegraphics[width=\textwidth]{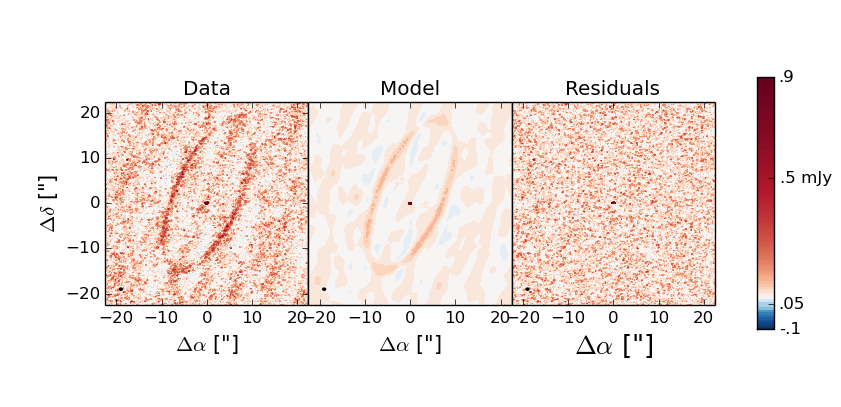}
\caption{\textbf{Left:} CLEANed data of the Fomalhaut system. The synthetic beam is given by the black ellipse in the bottom left of the image. \textbf{Middle:} MCMC constrained best fit model of the system and simulated in {\scriptsize CASA} by predicting onto the data visibilities. Resulting image is CLEANed with the same mask as the actual data. \textbf{Right:} The data minus model residuals from the best fit model. The residuals are consistent with $\sim1.5~\sigma_{\rm RMS}$ of the observations. In all images North is up and East is to the left. The apparent excess in the centre of the image is an artefact due to gridding effects in generating the image.
\label{vis_resid}}
\end{figure*}

The data and best fit model are then deconvolved and imaged using the {\scriptsize CASA} CLEAN algorithm. The data, model, and residuals are shown in Fig. \ref{vis_resid}.

\begin{table}
\caption{Summary of visibility plane MCMC Results with $95\%$ Credible Range. As before, the width is the FWHM of the ring.} 
\centering 
\begin{tabular}{c | c | c} 
\hline\hline 

	Parameter & Most Probable & $95\%$ Credible Range \\
	\hline
	Centre [au]     & $139$  & $[136, 141]$ \\
	Width [au]   & $13$ & $[10, 16]$ \\
	Inclination [$^{\circ}$]   & $66.7$ & $[66.0, 67.2]$   \\
	Position Angle [$^{\circ}$]& $336.5$  & $[337.1, 335.9]$     \\
	X Offset [arcsec] & $-0.23$ & $[-0.35, -0.10]$ \\
    Y Offset [arcsec] & $-1.87$ &  $[-2.08, -1.64]$ \\
	Disc Flux [mJy] & $30.8$ & $[27.8, 34.2]$ \\
	Central Emission [mJy] & $0.90$ & $[0.83, 0.95]$ \

\label{mcmc_vis_par}
\end{tabular}
\end{table}

\subsection{Comparison between approaches}

The image-plane and visibility fitting methods are both very consistent with each other in describing the disc geometry.  The biggest discrepancy among the results of the two methods is the most probable flux for the debris ring, which has a difference of about $15\%$. In contrast, the flux for the central emission is only $\sim 1\%$ different and well within the uncertainties at the 95\% confidence level. If observations are simulated in {\scriptsize CASA} using  the best fit image-plane model, the resulting residuals reveal a small amount of leftover flux in the location of the ring. Furthermore,  we cautiously note that the flux derived by fitting the visibilities is the most consistent with expectations based on extrapolating the 870 $\mu$m results. Visibility fitting appears to be  the most accurate approach.

Nonetheless, fitting in the image plane can still be advantageous. In particular, the MCMC quickly converged on the debris disc morphology; most of the best fit geometrical parameters are well within the $1\sigma$ results of those derived from  fitting to the visibilities.  If only the disc geometry is needed,  then image-plane fitting could be a reasonable approach, as advocated by \citet{booth}.  Furthermore,  a preliminary best model could be selected by fitting in the image-plane, particularly as using poor starting conditions can have a major impact on the required links for an MCMC to converge and fitting visibilities can be time-consuming.  In some cases, fitting first in the image plane and then refining the fit using the visibilities maybe improve model selection.

\subsection{Additional Properties of the Debris System}

The Gaussian profile of the chosen model accurately recovers the geometry of the disc, as can be seen in the residuals in Fig. \ref{vis_resid}. Power law disc models were considered, but were not well constrained in the model fitting. As noted above, the ring ansae are close to the edges of the primary beam. As such, any potential North/South asymmetries are not reliable. Furthermore, there are no noted deviations from an azimuthally smooth ring to within the noise level of the measurements. The ``fading out" of the ring and then brightening at the ansae is an artefact of the primary and synthetic beams and is reproduced  with the azimuthally symmetric models (see middle panels of Fig.\,\ref{img_resid} and Fig.\,\ref{vis_resid}).

Other than the unresolved central emission, there is no detection of any structure or emission interior or exterior to the debris ring. Assuming that the debris ring has only a small eccentricity with the star at one of the foci, the ring's eccentricity can be calculated from the best fit semi-major axis and the X and Y offsets of the ellipse centre, as derived from the most probable model values listed in Table \ref{mcmc_vis_par}. Thus,
\begin{equation}
e = \frac{({\rm X'^2 + Y'^2})^{1/2}}{a},
\end{equation}
which yields an eccentricity of $0.12 \pm0.03$ with uncertainties propagated from the $95\%$ Credible Ranges listed in Table \ref{mcmc_vis_par}. X' and Y' are the de-projected offsets, i.e., they represent the offsets if the system were viewed face-on, before the disc is inclined and rotated by the PA. The eccentricity result is in agreement with the e $=0.11\pm0.01$ from \textit{HST} scattered light observations \citep{kalas}. 

This measured eccentricity is relatively high and has implications for potential disc-planet interactions within the Fomalhaut system. One likely scenario is that the ring has a forced eccentricity due to an interior massive planet \citep[e.g., see][]{wyatt99, kalas}.  An inner planetary system could also give rise to a sharp inner edge, but additional dynamics may be required to explain the abrupt outer edge as well  \citep{boley}.  While the main Fomalhaut ring has a large forced eccentricity, the millimetre grains are narrowly located within a radial region that has a FWHM of 13 au. Because the millimetre grains are not strongly affected by radiation pressure, these grains further suggest that the collisional parent body population is also narrowly located.

A rough mass estimate can also be made by making a few simplifying approximations for the disc. We assume that the debris is comprised of 1.3 mm grains, i.e., the wavelength of the observations, and that they are perfect radiators in thermal equilibrium with the host star.   All of the grains are placed at a distance 139 au from the star.  Adopting an average density of 2.5$~g~cc^{-1}$, this approach yields a mass of M$_{\rm 1.3\,mm}\sim 0.017~M_{\bigoplus}$. This is in agreement with the simple mass \citet{boley} derived for the ALMA 345 GHz observations and can be interpreted as a lower limit to within the assumed density of the grains.  

The above simple mass calculation is incomplete in that it does not consider how a distribution of grain sizes, up to some parent body size, can affect the total debris mass.  To illustrate this, we use the method laid out in \cite{white} to estimate the debris mass contained within objects of a given grain size distribution. The grains are assumed to radiate efficiently as long as the their circumference is equal to or larger than the absorbing/emitted photons \citep{draine}. For wavelengths larger than the grain's circumference\footnote{If the gain's diameter is used  instead \citep[e.g., see][]{wyatt}, then this removes a factor of $\pi$ from the absorption coefficient. This will affect the total mass by a factor $\sim2$, yielding $\sim11\rm~M_{\bigoplus}$ instead of $\sim6\rm~M_{\bigoplus}$ as given in Eq. 7.}, the emission and absorption coefficients are inversely proportional to the photon wavelength. The flux density for any given grain is calculated by assuming that the albedo $A\sim0$ and that the received and emitted powers balance, using a black body model modified to take into account the emission and absorption coefficients.  The relative flux density for each bin of grain sizes is then evaluated.   The total mass is then determined by requiring that the flux density of the model match the flux density derived from the observations (30.8 mJy in this case).  We only consider a power law size distribution characterized by $q=3.5$ for $\frac{dN}{dD}\propto D^{-q}$, where $D$ is the grain or planetesimal diameter.  The chosen value for $q$ is consistent with a collisional cascade, and as will be shown below, is a reasonable estimate for the millimetre grains in Fomalhaut's debris ring (see Section 4).   

The total estimated mass will be heavily influenced by the maximum grain size. In our own Solar System, the Kuiper Belt has a strong drop off in material, or ``knee", for objects with a diameter of 50 km \citep{gladman}. If we further adopt a Kuiper Belt like grain density of 1$~g~cc^{-1}$, which is more appropriate for cometary-like material, we find a total mass of M$(\rm D<50\,km) \sim 6~ M_{\bigoplus}$. This assumes that the timescale for collisions with $D\sim50$ km-sized objects is short enough for these objects to contribute to the cascade.  While is is unclear whether this applies to Fomalhaut, it is a working assumption for comparison with the Kuiper Belt. A more general mass relation for the collisional cascade can be written  as
\begin{equation}
M(<D) \approx 6 ~ \frac{\rho}{1~g~cc^{-1}} ~\Bigg(\frac{D}{50~km}\Bigg)^{\frac{1}{2}} \rm M_{\bigoplus}
\end{equation}
for a given maximum diameter, $D$, and density, $\rho$.

The Kuiper Belt has a total inferred mass for $\rm D<50\,km$ of $\sim0.1~M_{\bigoplus}$ \citep{gladman}. This means that there is potentially 60 times more collisional material in the Fomalhaut debris ring than in the current Kuiper Belt, assuming the size cutoffs are appropriate. It should be cautioned though that the current mass of the Kuiper Belt is likely smaller than it was when our Solar System was the same age as Fomalhaut. Estimating the amount of mass that the Kuiper Belt has lost due to collisional erosion over its lifetime may be model dependent.  In the Kuiper Belt, the main mass loss mechanism is the dynamical ``erosion'' of the scattering population due to gravitational interactions with planets \citep[e.g., see][and references therein]{lawler}.

\section{SED Modelling}

As noted above, the estimate for the debris ring's mass is dependent on the size distribution of grains. The grain size distribution can be inferred from the slope of the flux density, assuming $F_\nu\propto \nu^{\alpha_{\rm mm}}$, where $\alpha_{\rm mm}$ is the spectral index at millimetre wavelengths. We calculate $\alpha_{\rm mm}$ by combining the ring's 1.3 mm flux density (30.8 mJy, as derived from the visibility fitting here) with literature values for flux densities at different wavelengths (see Table \ref{flux_list}). The posterior distribution for the spectral index is determined by performing a Bayesian parameter estimation for $F_\nu$. We use an MCMC approach similar to that used for the disc model fitting, incorporating the listed flux uncertainties, and assume a flat prior distribution. We fit over the wavelengths 350 - 1300 $\mu$m and 350 - 6600 $\mu$m separately, yielding most probable values of $\alpha_{\rm mm}=-2.62 \pm 0.12$ and $\alpha_{mm} = -2.73 \pm 0.13$, respectively. The $1\sigma$ uncertainties are also given.

The slope of the grain size distribution, $q$, for $\frac{dn}{ds}\propto s^{-q}$,  is given by the following \citep[see e.g.][]{dalessio, ricci, macgregor}:
\begin{equation}
q = \frac{\alpha_{mm} - \alpha_{pl}}{\beta_{s}} + 3,
\end{equation}
where $\alpha_{pl}$ is the spectral index of the Planck function over the wavelengths of interest, and $\beta_{s}$ is the dust opacity spectral index in the Rayleigh limit. Following \citet{ricci}, we adopt $\alpha_{pl} = 1.84 \pm 0.02$ and $\beta_{s} = 1.8 \pm 0.2$.

The adopted value of $\alpha_{pl}$ would normally be 2 in the Rayleigh-Jeans limit. However, the actual value of $\alpha_{pl}$ depends on the temperature of the dust and the wavelengths of interest \citep[e.g,][]{holland03}.  Specifically,
\begin{equation}
\alpha_{pl} = \left| \frac{\rm log\Big(\frac{B_{\nu_{1}}}{B_{\nu_{2}}}\Big)}{\rm log\Big(\frac{\nu_{1}}{\nu_{2}}\Big)}\right|,
\end{equation}
where B$_{\nu}$ is the Planck Function and $\nu_{1}$ and $\nu_{2}$ are, e.g., the respective frequencies for the 350 and 6600 $\mu$m observations.
The derived temperature range for Fomalhaut's ring is approximately between 40 - 50 K.  Assuming a dust temperature of $45\pm5$ K yields $\alpha_{pl} = 1.84 \pm 0.02$, as used in \cite{ricci}.
Finally, \citet{draine} find that for particles larger than $100~\mu$m, $\beta_{s}$ in discs is consistent with the grains in the ISM, which means $\beta_{s} \approx  \beta_{ism} = 1.8 \pm 0.2$, as long as $3 < q < 4$. It is worth noting though that the value of $\beta_{s}$ could range between 1.0 and 2.0. $\beta_{s}\approx 1.0$ has been found in protoplanetary discs \citep{andrews} and $\beta_{s}=2.0$ is found in simple models of conductors/insulators \citep{draine04}. Adopting different values of $\beta_{s}$ can lead to significantly different values of q.

Using $\beta_s=\beta_{ism}$, the calculated grain size distribution with the addition of the ALMA band 6 observations is $q = 3.50\pm 0.14$, consistent with the previously calculated values of $q = 3.48 \pm 0.14$ \citep{ricci}. Using the 350 - 1300 $\mu$m data set gives $q = 3.43\pm 0.15$, which is a bit shallower, but still consistent within the $1\sigma$ uncertainties. Both of these results are also consistent with the predicted $q=3.51$, which would be expected for a steady-state collisional cascade model \citep{dohnanyi}, similar to that used in our mass estimates for the ring. Strictly, this value reflects the size distribution for approximately millimetre grains, and does not necessarily extend to other size regimes.

\begin{table*}
\caption{List of select, previous observations of Fomalhaut from the literature. The uncertainties, when not listed, are assumed to be 10\%. The (*) denotes the ALMA 870 $\mu$m observations by \citet{boley}. The central emission was located near the edge of the primary beam and as such the flux estimate is not as reliable as the flux from \citet{su16}, where the central emission was at the phase centre of the observations. The $870~\mu$m flux value from \citet{su16} was used in all analysis.} 

\centering 
\begin{tabular}{c | c | c | c | c | c} 
\hline\hline 

    Wavelength [$\mu$m] & Disc Flux [mJy] & Uncertainty [mJy] & Star Flux [mJy] & Uncertainty [mJy]  & Reference \\
	\hline
		6600  & 0.308  & - & 0.092  & $\pm 0.015$  & \citet{ricci} \\
		1300  & 30.8  & $\pm 4.1$  & 0.89  & $\pm 0.149$  & This Work\\
		870  & 85  & $\pm 8.5$ & 3.4$^{*}$  & $\pm 0.34$  & \citet{boley}\\
		870  & - & - & 1.79  & $\pm 0.216$ & \cite{su16}\\
		850  & 97  & - &  - & - & \cite{holland03}\\
		850  & 81  & - &  - & - & \cite{holland98}\\
		500  & 345  & $\pm 35$  & 10  & - & \cite{acke} \\
		450  & 595  & $\pm 35$ & -  & - & \cite{holland03}\\
		350  & 595  & $\pm 35$ & 22  & - & \cite{acke}\\
		250  & 1970  & $\pm220$   & 54  & - & \cite{acke}\\
		160  & 4650  & $\pm450$   & 124  & - & \cite{acke}\\
		70  & 7990  & $\pm666$   & 540  & - & \cite{acke}\\
		\hline
		Wavelength [$\mu$m] & Disc Flux [Jy] & Uncertainty [Jy] & Star Flux [Jy] & Uncertainty [Jy]  & Reference \\
		\hline
		24  & - & - & 2.96  & $\pm 0.29$  & \cite{su16}\\
		18.4  & - & - & 5.34  & $\pm 0.08$  & \cite{ishihara} \\
		8.6  & - & - & 23.0  & $\pm 0.04$  & \cite{ishihara}\\
		2.16  & - & - & 257 & - & \cite{pickles}\\
		1.65  & - & - & 399 & - & \cite{pickles}\\
		1.24  & - & - & 594 &- & \cite{pickles}\\
		0.554  & - & - & 1250 & - & \cite{boyajian}\

\label{flux_list}
\end{tabular}
\end{table*}

\subsection{What can the observations tell us about a possible close in warm debris system?}

There is a clear detection of the central emission at the phase centre of the images (see Figs. \ref{img_resid}, \ref{vis_resid}). The best fit flux from the MCMC visibility modelling is 0.90$\pm 0.15$ mJy. The photosphere temperature of the star is constrained to be T$_{B} = 8600 \pm200$ K from \textit{Herschel} observations \citep{acke}.  Assuming that this brightness temperature also reflects the flux density at longer wavelengths, we would naively expect the  1300 $\mu$m flux density to be 1.3 mJy. The ALMA 1300 $\mu$m observations recover  $<70\%$ of the this flux.  This is consistent with the the 870 $\mu$m observations, where \cite{su16} recover 1.8 mJy  ($\sim 2.8$ mJy is expected based on a black body with the stellar photosphere temperature).  Even if the measured flux density in the far-infrared were to be extrapolated to millimetre wavelengths, the millimetre flux density would still be lower than expected.

To determine the presence of an inner debris component, there first needs to be an accurate characterization of the flux contribution of the host star. Because the observed flux is already lower than expected, this means that the brightness temperature of the Fomalhaut star  at these wavelengths is much less than that of the photosphere and even less than in the far-infrared (although consistent within the uncertainties). As such, the degree of unresolved excess emission in the inner system cannot be easily determined. 

\begin{figure}
\centering
\includegraphics[width=.5\textwidth]{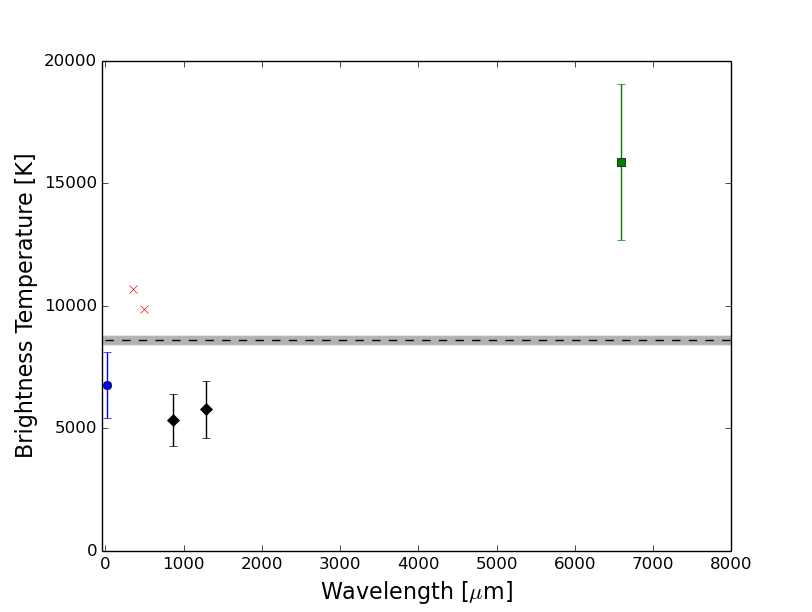}
\caption{Brightness temperature of the star from the recovered flux at a given wavelength. The horizontal dashed line represents a brightness temperature equal the stellar photosphere of T$_{B} = 8600 \pm200$ K with the grey region representing the uncertainty. The blue circle is the 24 $\mu$m data, the black diamonds are ALMA data, and the green square is ATCA data. The \textit{Herschel} data are denoted as X's since they are not direct measurements of the star. For flux values and uncertainties see table \ref{flux_list}. 
\label{TB_plot}}
\end{figure}

If it is assumed that there is no inner debris component and all the observed flux is intrinsic to the star, then processes in the stellar atmosphere (e.g. chromospheric opacity effects) must be causing significant changes to the brightness temperature of the star at millimetre wavelengths.  Fig. \ref{TB_plot} shows the 24 $\mu$m and the 350-6600 $\mu$m recovered fluxes along with the the brightness temperatures that the star would need to have to produce each measurement. The horizontal dashed line represents a brightness temperature equal to the stellar photosphere of T$_{B} = 8600 \pm200$ K.  Around wavelengths of 1 mm, the brightness temperature drops below 65\% of the photosphere temperature. At larger wavelengths the brightness temperature increases to nearly double that of the photosphere temperature. This is very similar to the behaviour observed in the Sun at mm/submm wavelengths (e.g., Fig. 1 in \citet{loukitcheva}), in which the observed solar flux drops down to $\sim80\%$ of the Sun's photosphere brightness temperature before increasing with increasing wavelength.  This profile for the millimetre flux densities ultimately reflects different layers in the chromosphere, with longer wavelengths probing higher atmosphere altitudes  \citep[e.g., see][]{wedemeyer}. ALMA 440-3100 $\mu$m observations of the $\alpha$ Centauri system \citep{liseau} also find Solar chromosphere-like behaviour in the binary. The G2V and K1V star's stellar atmospheres indicate that the observed trends in the brightness temperature are not exclusive to the Sun. The Fomalhaut observations allow us to begin to explore such behaviour in an A star, assuming any inner dust is negligible.

Using the stellar fluxes given in Table \ref{flux_list}, we fit black bodies to subsets of the flux density data using a Bayesian approach. In the ALMA 870 $\mu$m observations by \citet{boley} the central emission was located near the edge of the primary beam. As such, flux estimate is not as reliable as the flux from \citet{su16}, where the central emission was at the phase centre of the observations. The $870~\mu$m flux value from \citet{su16} was used in all further analysis.

The corresponding brightness temperatures for specific wavelength ranges are given in Table \ref{temp_list}. Fitting to all the stellar data, 0.554 - 6600 $\mu$m, as well as the wavelengths 0.554 - 24 $\mu$m, yields T$_{B} = 8650$ K. This value is very much in line with the previously constrained photosphere temperature of T$_{B} = 8600 \pm200$ K \cite{acke}. If instead only the 870-6600 $\mu$m data are used, then the expected brightness temperature is T$_{B} = 5540$ K.  This value is less than 65\% of what can be ``expected" from assuming the brightness temperature is the same as the stellar photosphere temperature. 

\begin{figure*}
\centering
\includegraphics[width=\textwidth]{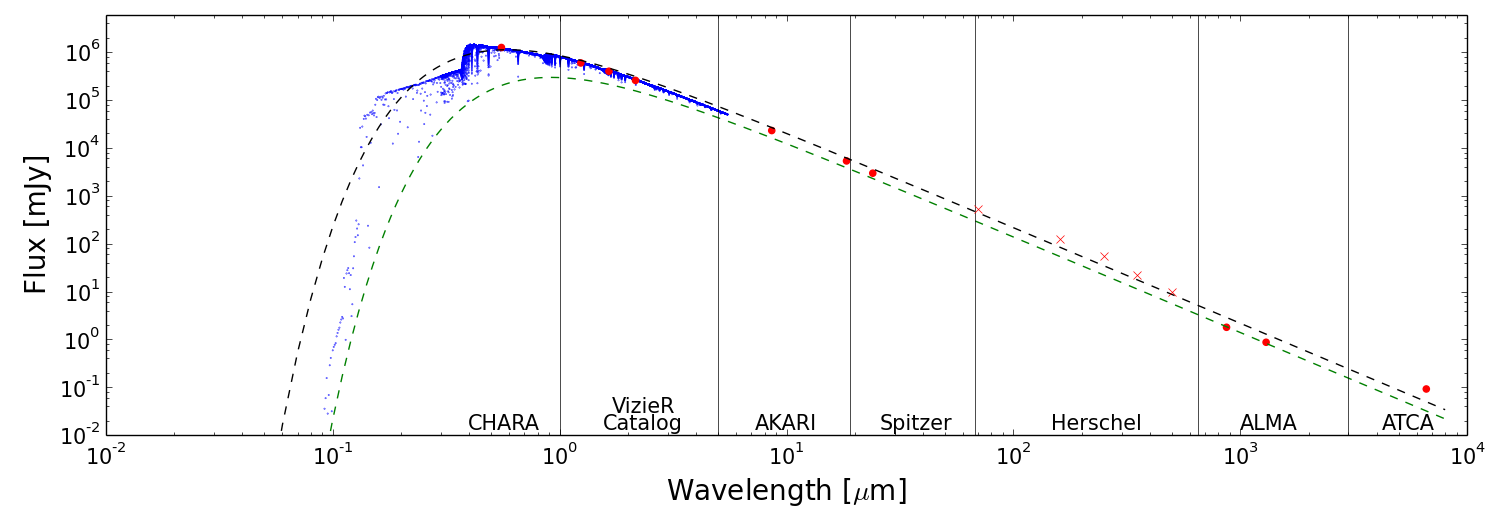}
\caption{Data from literature and the corresponding fits to a black body. The blue dots represents a PHOENIX stellar atmosphere similar to Fomalhaut \citep{husser}. The black dashed line is the best fit black body for all data points with T$_{B} = 8647$ K. The green dashed line is the 870 $\mu$m - 6600 $\mu$m data with T$_{B} = 5540$ K. The \textit{Herschel} data are denoted as red X's because they are not direct measurements of the star. The vertical lines represent different regions measured by different observatories.
\label{BB_plot}}
\end{figure*}

Fig.~\ref{BB_plot} shows all of the stellar flux data from Table \ref{flux_list}, along with two black bodies with T$_B=8600$ K (the photosphere) and T$_B=5540$ K (from millimetre data).  A PHOENIX Stellar atmosphere model similar to that of Fomalhaut \citep{husser} is also shown. The \textit{Herschel} data from 70-500 $\mu$m are not a direct measurement of the stellar emission, but instead are the estimated stellar contribution to the unresolved central emission at each corresponding wavelength. Thus, these data points may not accurately represent the stellar emission at far-infrared/submm wavelengths. Overall, the emission centred on the star Fomalhaut does not show clear evidence for an inner debris system.  However, precise limits on any excess emission over the stellar emission, should it exist, can only be done if the stellar emission of A stars is properly characterized at submillimetre and millimetre wavelengths.  Dynamical processes in the stellar atmosphere may further be a source of significant deviations in brightness temperature at submm/mm wavelengths \citep{wedemeyer}, further confounding the problem.

\begin{table}
\caption{List of data sets selected for SED fitting. A black body was fit to each data subset through a Bayesian approach that includes the uncertainties. For data from literature, when the uncertainty is not given, a 10\% uncertainty is used. The best fit brightness temperature and 95\% credible region are given. The (*) denotes the range of data in the \textit{Herschel} observations. As these are inferred values, and not direct measurements, for the star, they may not accurately represent the stellar flux at these wavelengths.}

\centering 
\begin{tabular}{c | c | c  } 
\hline\hline 

    Data Range [$\mu$m] & Brightness Temp [K] & Uncertainty [K]  \\
	\hline
		0.554  - 6600 & 8647 & 8645-8649  \\
		0.554  - 2.16  & 8651 & 8650-8652  \\
		2.16  - 24  & 7611 & 7605-7617   \\
		70 - 6600 & 10,750$^{*}$ & 10,500-11,020 \\
		870  - 6600  & 5540 & 3570-7860   \\
		870  - 1300  & 5550 & 3530-7910  \

\label{temp_list}
\end{tabular}
\end{table}

\section{Summary}

We have presented ALMA band 6 observations of the Fomalhaut debris system. These 0.28 arcsec resolution observations targeted both the outer debris ring as well as the central emission around the host star. A two component model was fit to the data, consisting of a a ring with a Gaussian radial width and a point source for the central emission. The best fit model recovered a flux of 30.8 mJy for a ring centred at 139 au and a FWHM of 13 au.  The system inclination was found to be 66.7$^{\circ}$ with a position angle of 336.5$^{\circ}$.  The best fit model's ring has a projected  X, Y offset of -0.23 arcsec and -1.87 arcsec from the central emission, which was found to have a flux density of 0.90 mJy. Model fitting was conducted using the visibilities and the image-plane separately, and while the image-plane was able to consistently recover the geometry and central emission, we find that there is a $\sim15\%$ discrepancy in the amount of recovered ring flux. We conclude that visibility fitting remains necessary, but image-plane fitting can be used to determine preliminary models.  

The spectral index of the mm grains within Fomalhaut's debris ring was constrained to be $\alpha_{mm} = -2.64 \pm 0.12$ for wavelengths from 350 - 1300 $\mu$m, and $\alpha_{mm} = -2.73 \pm 0.13$ for 350 - 6600 $\mu$m. This corresponds to a grain size distribution of $q = 3.43\pm 0.15$ and $q = 3.50\pm 0.14$, respectively, consistent with a steady state collisional cascade model.

The 0.28 arcsec resolution of the observations is about 2.1 au at the distance of the system. There is no detected extended structure or any obvious excess emission over the intrinsic stellar flux.  Instead, we find that the fitted 0.90 mJy of flux density corresponds to a stellar brightness temperature of 5540 K, less than 70\% of what can be expected by assuming the millimetre brightness temperature is the same as the stellar photosphere temperature. This is likely due to the star's chromosphere, analogous to the Sun. The ALMA observations of the Fomalhaut star presented here are part of an ongoing project in measuring the emission of stellar atmospheres at submm/mm wavelengths.

\section*{Acknowledgements}

We thank the anonymous referee for the useful feedback. The authors would like to acknowledge Matthew J. Payne for help in obtaining this dataset. J.A.W. and A.C.B acknowledge support from an NSERC Discovery Grant, the Canadian Foundation for Innovation, The University of British Columbia, and the European Research Council (agreement number 320620).  E.B.F. acknowledges support from the Center for Exoplanets and Habitable Worlds.  The Center for Exoplanets and Habitable Worlds is supported by the Pennsylvania State University, the Eberly College of Science, and the Pennsylvania Space Grant Consortium. 

This paper makes use of the following ALMA data: ADS/JAO.ALMA[2013.1.00486.S] . ALMA is a partnership of ESO (representing its member states), NSF (USA) and NINS (Japan), together with NRC (Canada), NSC and ASIAA (Taiwan), and KASI (Republic of Korea), in cooperation with the Republic of Chile. The Joint ALMA Observatory is operated by ESO, AUI/NRAO and NAOJ





\begin{thebibliography}{99}
\bibitem[Absil et al.(2009)]{absil}Absil, O., Mennesson, B., Le Bouquin, J.B., et al. 2009, ApJ, 704, 150

\bibitem[Acke et al.(2012)]{acke}Ackem B., Min M., Dominik C., et al., 2012, A\&A, 540, A125

\bibitem[Andrew \& Williams (2005)]{andrews}Andrews S.M. \& Williams J.P., 2005, ApJ, 631, 1134

\bibitem[Boley et al.(2012)]{boley}Boley A.C., Payne M.J., Corder S., et al., 2012, ApJL 750(1), p.L21

\bibitem[Booth et al.(2016)]{booth}Booth M., Jord\'an A., Casassus S., et al., 2016, MNRAS 460(1), pp.L10-L14

\bibitem[Boyajian et al.(2013)]{boyajian}Boyajian T.S., von Braun K., van Belle G., et al., 2013, \apj, 771(1), p.40

\bibitem[D'Alessio et al.(2001)]{dalessio}D'Alessio P., Calvet N., Hartmann, L. 2001, ApJ, 553, 321

\bibitem[Di Folco et al.(2004)]{difolco}Di Folco E., Th\'evenin F., Kervella P., et al. 2004, A\&A, 426, 601 

\bibitem[Dohnanyi(1969)]{dohnanyi}Dohnanyi B.T., 1969, J. Geophys. Res., 74, 2531

\bibitem[Draine(2004)]{draine04}Draine B.T., 2003, ApJ, 598, 1026

\bibitem[Draine(2006)]{draine}Draine B.T., 2006, ApJ, 636, 1114

\bibitem[Ford(2005)]{ford}Ford E.B., 2005, AJ, 129(3), 1706

\bibitem[Gladman et al.(2001)]{gladman}Gladman B., Kavelaars J.J., Petit J.M., et al., 2001, \aj, 122(2), p.1051

\bibitem[Holland et al.(2003)]{holland03}Holland W.S., Greaves J.S., Dent W.R.F., et al., 2003, \apj, 582(2), p.1141

\bibitem[Holland et al.(1998)]{holland98}Holland W.S., Greaves J.S., Zuckerman B., et al., 1998, Nature, 392(6678), pp.788-791

\bibitem[Husser et al.(2013)]{husser} Husser T.O., Wende-von Berg S., Dreizler S., et al., 2013, A\&A, 553, p.A6

\bibitem[Ishihara et al.(2010)]{ishihara}Ishihara D., Onaka T., Kataza H., et al., 2010, A\&A, 514, p.A1

\bibitem[Janson et al.(2012)]{janson}Janson M., Carson J.C., Lafreni\'ere D., et al. 2012, ApJ, 747, 116

\bibitem[Kalas et al.(2005)]{kalas}Kalas P., Graham J.R., \& Clampin M., 2005, Nature, 435, 1067

\bibitem[Kalas et al.(2008)]{kalas2008}Kalas P., Graham J.R., Chiang E., et al. 2008, Science, 322, 1345

\bibitem[Kavelaars et al.(2008)]{kavelaars}Kavelaars J.J., Jones L., Gladman B., et al., 2008, The Solar System Beyond Neptune, pp.59-69

\bibitem[Lawer et al.(2015)]{lawler}Lawler S.M., Greenstreet S., \& Gladman, B., 2015, ApJL, 802(2), p.L20

\bibitem[Liseau et al.(2016)]{liseau} Liseau R., De la Luz V., O'Gorman E., et al., 2016, arXiv preprint arXiv:1608.02384.

\bibitem[Loukitcheva et al.(2004)]{loukitcheva} Loukitcheva M., Solanki S.K., Carlsson M., Stein R.F., 2004, A\&A, 419(2), p.747-756

\bibitem[Lyra \& Kuchner(2012)]{lyra}Lyra W. \& Kuchner M., 2013, \nat, 499(7457), pp.184-187

\bibitem[MacGregor et al.(2016)]{macgregor} MacGregor M.A., Wilner D.J., Chandler, C., et al., 2016, ApJ, 823(2), p.79

\bibitem[Mamajek(2012)]{mamajek}Mamajek E.E. 2012, ApJL, 754, L20

\bibitem[Marengo et al.(2009)]{marengo}Marengo M., Stapelfeldt K., Werner M. W., et al. 2009, ApJ, 700, 1647

\bibitem[McMullin et al.(2007)]{casa_reference}McMullin J.P., Waters B., Schiebel D., et al., 2007, Astronomical Data Analysis Software and Systems XVI (ASP Conf. Ser. 376), ed. R. A. Shaw, F. Hill, \& D. J. Bell (San Francisco, CA: ASP), 127

\bibitem[Mustill \& Wyatt(2008)]{mustill} Mustill A.J., Wyatt M.C., 2009, MNRAS, 399, 1403

\bibitem[Pan \& Schlicting(2012)]{pan}Pan M., Schlichting H.E., 2012, \apj, 747(2), p.113

\bibitem[Pickles \& Depagne(2010)]{pickles}Pickles A., Depagne \'E., 2010, Pubs. of the Astr. Soc. of the Pacific, 122(898), p.1437

\bibitem[Quillen(2006)]{quillen2006}Quillen A.C. 2006, MNRAS, 372, 14

\bibitem[Ricci et al.(2012)]{ricci}Ricci L., Testi L., Maddison S.T., Wilner D.J., 2012, A\&A, 539, p.L6

\bibitem[Stapelfeldt et al.(2004)]{stapelfeldt}Stapelfeldt K.R., Holmes E.K., Chen C., et al., 2004, ApJS, 154, 458

\bibitem[Su et al.(2016)]{su16}Su K.Y., Rieke G.H., Defr\'ere D., et al., 2016, \apj, 818(1), p.45

\bibitem[van Leeuwen(2007)]{hipparcos} Van Leeuwen F., 2007, A\&A, 474(2), 653

\bibitem[Vandenbussche et al.(2010)]{vandenbussche}Vandenbussche B., Sibthorpe B., Acke B.,  et al., 2010, A\&A, 518L, 133

\bibitem[Wedemeyer et al.(2015)]{wedemeyer}Wedemeyer S., Bastian T., Brajsa R., et al., 2015, ArXiv e-prints 1504.06887

\bibitem[White et al.(2016)]{white} White J.A., Boley A.C., Hughes A.M., et al., 2016, \apj, 829(6), p.11

\bibitem[Wyatt et al.(1999)]{wyatt99} Wyatt M.C., Dermott S.F., Telesco C.M., et al., 1999, ApJ, 527, 918

\bibitem[Wyatt \& Dent (2002)]{wyatt}Wyatt M.C. \& Dent W.R.F., 2002, MNRAS, 334, 589

\bibitem[Wyatt(2008)]{wyatt08} Wyatt M.C., 2008, ARA\&A, 46, 339
\end{thebibliography}




%
%


\bsp	
\label{lastpage}
\end{document}